\documentclass[aps,pra,twocolumn,groupedaddress,showpacs,floatfix,superscriptaddress]{revtex4-2}

\usepackage[T1]{fontenc}
\usepackage[latin9]{inputenc}
\usepackage{natbib}
\usepackage{babel}
\usepackage{amsmath}
\usepackage{amssymb}
\usepackage{graphicx}
\usepackage{enumitem}
\usepackage[unicode=true, bookmarks=false, breaklinks=false,pdfborder={0 0 1},backref=false,colorlinks=false]
 {hyperref}

\usepackage{breakurl}

\makeatletter

 \@ifundefined{textcolor}{}
 {
   \definecolor{BLACK}{gray}{0}
   \definecolor{WHITE}{gray}{1}
   \definecolor{RED}{rgb}{1,0,0}
   \definecolor{GREEN}{rgb}{0,1,0}
   \definecolor{BLUE}{rgb}{0,0,1}
   \definecolor{CYAN}{cmyk}{1,0,0,0}
   \definecolor{MAGENTA}{cmyk}{0,1,0,0}
   \definecolor{YELLOW}{cmyk}{0,0,1,0}
 }
\usepackage{bbm, dsfont}
\usepackage{bm}
\usepackage{amsfonts}
\hypersetup{colorlinks=true, citecolor=blue,linkcolor=blue, filecolor=blue, urlcolor=blue}

\newcommand{\ket}[1]{\left| #1 \right>} 
\newcommand{\bra}[1]{\left< #1 \right|}

\makeatother
\newcommand{\abs}[1]{\left| #1 \right|}

\begin{document}

\title{Ergotropy-Based Quantum Thermodynamics}

\author{J. M. Z. Choquehuanca}
\affiliation{Instituto de F\'isica, Universidade Federal Fluminense, Av. Gal. Milton Tavares de Souza s/n, Gragoat\'a, 24210-346, Niter\'oi, RJ, Brazil}

\author{P. A. C. Obando}
\affiliation{School of Physics, University of the Witwatersrand, 1 Jan Smuts Avenue, Braamfontein, 2000, South Africa}

\author{M. S. Sarandy}
\affiliation{Instituto de F\'isica, Universidade Federal Fluminense, Av. Gal. Milton Tavares de Souza s/n, Gragoat\'a, 24210-346, Niter\'oi, RJ, Brazil}

\author{F. M. de Paula}
\affiliation{Centro de Ci\^ encias Naturais e Humanas, Universidade Federal do ABC, Avenida dos Estados 5001, 09210-580, Santo Andr\'e, S\~ao Paulo, Brazil}
\date{\today}
\begin{abstract}
We introduce an ergotropy-based formulation of quantum thermodynamics, which provides a strong connection between average heat and von Neumann entropy. By adopting this formulation, we can reinterpret the infinitesimal average heat in terms of the infinitesimal change of the passive state associated with the density operator behind the quantum dynamics. Such as entropy, this leads to a heat concept that is invariant under passive state transformations. As an application, the average heat can be used as a general non-Markovianity measure for unital maps. Moreover, a positive-semidefinite temperature naturally emerges in an out-of-equilibrium ergotropy-based scenario. Concerning the infinitesimal average work, it arises as the infinitesimal variation of ergotropy, as well as an extra passive work contribution in the case of a time-dependent Hamiltonian. As illustrations, we consider the thermodynamics of a single-qubit open system in the cases of generalized amplitude-damping and phase-damping channels. 
\end{abstract}

\maketitle
\section{Introduction}

Quantum thermodynamics is at the heart of the energetics behind quantum technologies~\cite{Landauer:61,Bennett:82,Auffeves:2022}. As their classical counterparts, work extraction and heat exchange rule the efficiency of a quantum thermal machine~\cite{Gemmer:book,Deffner:book}. In turn, this imposes physical constraints for the working of quantum information devices (see, e.g., Ref.~\cite{Adesso:book}). More specifically, 
the energy flow in a quantum system is described by the quantum version of the first law of thermodynamics, which was originally established by Alicki~\cite{Alicki1979}. In this standard formulation, internal energy is defined by the expectation value of the underlying system Hamiltonian governing
the quantum dynamics. Concerning average work, it is associated with a controllable variation of the energy gaps, which is induced by an external parameter (e.g., a magnetic field) via a time-dependent Hamiltonian. On the other hand, by interacting with an external environment, the system may exchange heat with the environment, with average heat reflected as a variation in the energy level populations induced by a time-dependent density operator. Restrictions and potential applications of quantum devices will then naturally follow the interplay between heat and work, with real-world interactions governing energy extraction and entropy production in a general out-of-equilibrium quantum evolution~\cite{Landi2021}.

From a foundational point of view, even though the internal energy is essentially well established, there is a debate on the barrier between work and heat definitions, so that the energy balance is ensured. For instance, in Refs.~\cite{Alipour2022, Ahmadi:2023, Bertulio2020}, it has been proposed an alternative definition for the average heat in terms of the variation of the eigenvalues of the density operator (instead of the standard definition given by the variation of the whole density operator itself), which connects the definition of heat with the variation of the von Neumann entropy.  By adopting this so-called {\it{entropy-based}} formulation, the energy balance in the first law of thermodynamics requires average work to include an additional environment-induced component, which may be present even for constant Hamiltonians, as long as there is a system-environment interaction present throughout the quantum dynamics. 

Looking for a stronger connection between heat and entropy, we
propose here an {\it{ergotropy-based}} formulation for quantum thermodynamics. Ergotropy can be defined as the maximum energy
that can be extracted from a quantum state by a cyclic unitary transformation~\cite{Allahverdyan2004,Choquehuanca2024}. In this scenario, states that cannot provide work
are called passive states. By considering a general 
completely positive trace-preserving (CPTP) map, we then define 
the infinitesimal average heat flow through the infinitesimal 
change of the passive state in the process. 
Thus, average heat becomes closely related 
to the entropy variation, with a connection even 
stronger than that obtained by the entropy-based formulation. 
Indeed, such as entropy, it will follow that the ergotropy-based heat concept turns out to be   
invariant under a passive state transformation, {\it i.e.}, a map from the instantaneous density operator to its passive counterpart. 
As an application, the 
ergotropy-based average heat can be used as a measure of 
non-Markovianity for unital maps, generalizing the result of 
Ref.~\cite{Choquehuanca2023}. Moreover, we can also show that, 
in contrast with the previous formulations, this definition of 
heat implies a positive-semidefinite out-of-equilibrium qubit 
temperature for a general open-system dynamics. 
This differs from the results for the qubit temperature in both the standard and entropy-based formulations~\cite{Vallejo2020,Vallejo2021}.
 
Concerning the average infinitesimal work flow, it naturally arises as the infinitesimal variation of ergotropy, as well as an extra passive work contribution in the case of a time-dependent Hamiltonian. In particular, we will show that, 
for finite transformations and constant Hamiltonians, the 
ergotropy-based quantum thermodynamics reduces to the formalism 
introduced in Ref.~\cite{Binder2015}. As illustrations, we will consider 
Markovian and non-Markovian qubit evolutions for the generalized 
amplitude damping (GAD) and phase damping (PD) channels, comparing our results with 
the previous formulations and exploring some possible perspectives. 

\section{First law of quantum thermodynamics}

Consider an arbitrary quantum system described by a density operator 
\begin{equation}
\rho=\sum_{n} r_n\ket{r_n}\bra{r_n}
\end{equation}
and a Hamiltonian 
\begin{equation}
H=\sum_{n} \varepsilon_n\ket{\varepsilon_n}\bra{\varepsilon_n}
\end{equation}
such that $r_n\geq r_{n+1}$ and $\varepsilon_n\leq \varepsilon_{n+1}$. The ergotropy of the state $\rho$ is defined as the maximum amount of energy that can be extracted from quantum system via cyclic unitary operation \cite{Allahverdyan2004}, i.e.,
\begin{equation}\label{ergotropy}
\mathcal{E}(\rho)=\max_{V \in \mathcal{U}}\{U(\rho)-U(V\rho V^{\dagger})\}=U(\rho)-U(\rho_{\pi}),
\end{equation}
being 
\begin{equation}
U=\text{tr}[\rho H]   
\end{equation}
the internal energy of the system, $\mathcal{U}$ the set of all unitary transformations, and the optimized state 
\begin{equation}
\rho_{\pi}=\sum_n r_n\ket{\varepsilon_n}\bra{\varepsilon_n}   
\end{equation}
is known as passive state. From Eq.~(\ref{ergotropy}), we can write $dU(\rho)=dU(\rho_{\pi})+d\mathcal{E}(\rho)$, where $dU(\rho_{\pi})=\text{tr}[d\rho_{\pi} H]+\text{tr}[\rho_{\pi} dH]$. Thus,  we can establish a first law of quantum thermodynamics based on concept of ergotropy as 
\begin{equation}
dU=\delta Q+\delta W,
\end{equation}
with
\begin{equation} \label{heatwork}
\delta Q\equiv\text{tr}[d\rho_{\pi} H]
\end{equation}
and
\begin{equation}
\delta W\equiv\text{tr}[\rho_{\pi} dH]+d\mathcal{E}, 
\end{equation}
with $\delta$ denoting  inexact differential.
Note that the work $\delta W$ demands a time-dependent Hamiltonian or ergotropy variation. In particular, ergotropy can be decomposed into incoherent and coherent parts~\cite{Francica2020, Shi2022,Choquehuanca2024}, with the coherent contribution playing a potentially significant role in the performance of quantum thermal devices. On the other hand, the heat $\delta Q$ is invariant under unitary transformations and requires a change of the passive state. Consequently, $\delta Q$ depends on the change $dS$ in the von Neumann entropy 
\begin{equation}
S=\text{tr}[\rho \sigma],
\end{equation}
where $\sigma=-k_B\text{ln}\rho$ denotes the entropy operator. In fact, defining the passive part of an arbitrary functional $f(\rho)$ by
\begin{equation}
f_{\pi}(\rho)\equiv f(\rho_{\pi}), 
\end{equation}
we can write 
\begin{equation}
 \delta Q=\delta Q_{\pi}   
\end{equation}
and 
\begin{equation}
dS=dS_{\pi}=\text{tr}[d\rho_{\pi} \sigma_{\pi}]. 
\end{equation}
Note that both $\delta Q$ and $dS$ necessarily depend on $d \rho_{\pi}$, being quantities invariant under the passive transformation $\rho \to \rho_{\pi}$. 
Indeed, the quantity $\text{tr}[d\rho_{\pi} H]$ has been identified in Ref.~\cite{Niedenzu2018} as the fraction of the exchanged energy between a quantum system 
and a bath that necessarily causes an entropy change. In the ergotropy-based formulation, this passive energy is interpreted as total heat, in analogy with equilibrium classical thermodynamics. 
Since the $d\rho_\pi$ is associated with a change in the reduced density operator for the system, the expression for the ergotropy-based heat is applicable to a general system-bath interaction, including the cases of single and  multiple baths.

\section{Temperature}

Before discussing the second law of thermodynamics in the ergotropy-based formulation, let us proceed by introducing a closed expression for the temperature $\mathcal{T}$ of a quantum system. In a general nonequilibrium quantum system, $\mathcal{T}$ can be obtained by taking the partial derivative of the von Neumann entropy with respect to the internal energy~\cite{Alipour2021,Alipour2022}:
\begin{equation}\label{temperature}
\frac{1}{\mathcal{T}}\equiv \left(\frac{\partial S}{\partial U}\right)_{\{x_i\}_{i=2}^{d^2-1}}=\frac{\text{Cov}(H,\sigma)}{\text{Cov}(H,H)},
\end{equation}
where $\{x_i\}$ is a set of independent parameters kept constant in the partial derivative and 
\begin{equation}
\text{Cov}(X,Y)=\text{tr}[XY]/d-\text{tr}[X]\text{tr}[Y]/d^2   
\end{equation}
is the covariance between the operators $X$ and $Y$ evaluated with respect to the maximally mixed state ${\mathbbm{I}/d}$, with $d$ denoting the dimension of the associated Hilbert space. We take
\begin{equation}
 x_i = \text{tr}[\rho\, O_i],
\end{equation}
which represent the mean values of traceless
orthonormal observables $\{O_i\}_{i=2}^{d^2-1}$. This set can be made complete by adding 
\begin{equation}
O_0=\mathbbm{I}/\sqrt{d},
\end{equation}
which denotes the
normalized identity operator, and 
\begin{equation}
O_1=\frac{H-\text{tr}[H]\mathbbm{I}/d}{\sqrt{\text{Cov}(H,H)d}} ,   
\end{equation}
which is the operator associated with the Hamiltonian.  As $dU_{\pi} =\delta Q$ for a zero-work process and $dS_{\pi}=dS$, a definition of temperature, denoted by $T$, compatible with the relationship between heat and entropy in the ergotropy-based formulation  is given by the passive part of $\mathcal{T}$,
\begin{equation}\label{ptemperature}
T(\rho)\equiv\frac{\text{Cov}(H,H)}{\text{Cov}(H,\sigma_{\pi})}.
\end{equation}
Note that Eq.~(\ref{ptemperature}) is obtained from Eq.~(\ref{temperature}) by taking the passive transformation $\sigma \to \sigma_{\pi}$ over the entropy operator. Since $\sigma_{\pi}$ is a passive state, the  functional $T(\rho)$ satisfies the properties (see Ref.~\cite{Alipour2021}):
\vspace{0.5cm}
\begin{enumerate}[label=(\alph*), align=left, leftmargin=*, itemsep=0pt, topsep=0pt]
    \item Positivity:
    \begin{equation}
      T(\rho)\geq 0\,\,\, \forall\,\,\,\rho.  
    \end{equation}
    \item Nullity for pure states:
    \begin{equation}
      T(\ket{\psi}\bra{\psi})=0.  
    \end{equation}
    \item Divergence for maximally mixed states: 
    \begin{equation}
   T(\mathbbm{I}/d)\rightarrow\infty.
    \end{equation}
    \item Invariance under unitary operations:
    \begin{equation}
    T(V\rho V^{\dagger})=T(\rho). 
    \end{equation}
    \item Consistency with the Gibbsian state: 
    \begin{equation}
    T(\rho_G)=\frac{1}{k_B\beta}\,\,\,\,\,\,\text{with}\,\,\,\,\,\,\rho_G=\frac{e^{-\beta H}}{\text{tr}[e^{-\beta H}]}.
    \label{TGibbs}
    \end{equation}
\end{enumerate}
  By defining the inverse temperature as the partial derivative of the von Neumann entropy with respect to the internal energy under a zero-work process, we isolate the thermal contribution to changes in internal energy. Thus, temperature dictates how internal energy varies due to heat exchange. As described in the next section, this temperature also governs the irreversibility associated with entropy increase. Therefore, the notion of temperature introduced here serves as a conceptual bridge between the first and second laws of quantum thermodynamics within the ergotropy-based formulation.
\section{Second law of quantum thermodynamics}

Let us consider a CPTP evolution dictated by a dynamical map with a fixed point $\rho_*$, i.e.,
\begin{equation}
\rho(t)=\mathcal{M}(\rho_0) \,\,\,\,\,\text{such that }\,\,\,\,\,\mathcal{M}(\rho_*)=\rho_*.
\end{equation}
In this scenario, for an infinitesimal time interval $dt$, the entropy production is given by \cite{Landi2021}
\begin{equation}
 \delta\Sigma= S(\rho(t)||\rho_*)-  S(\rho(t+dt)||\rho_*) \equiv -\delta S(\rho||\rho_*)\geq 0, 
\end{equation}
where 
\begin{equation}
S(\rho||\rho_*)=k_B\text{tr}[\rho(\ln{\rho}-\ln{\rho}_*)] 
\end{equation}
provides the relative entropy between $\rho$ and $\rho_*$. This inequality is a direct consequence of the contractivity of the relative entropy under CPTP maps, i.e.,  $S(\mathcal{M}(\rho_0)||\mathcal{M}(\rho_*))\leq S(\rho_0||\rho_*)$. Similarly to the variation of internal energy, entropy production does not depend on the specific definitions of heat and work, exhibiting passive $\delta\Sigma_{\pi}$ and non-passive $\delta\Sigma_{n\pi}$ components:
\begin{equation}
\delta\Sigma=\delta\Sigma_{\pi}+\delta\Sigma_{n\pi}
\end{equation}
where
\begin{equation}
\delta\Sigma_{\pi}=-\delta S(\rho_{\pi}||\rho_*)
\end{equation}
and
\begin{equation}
\delta\Sigma_{n\pi}=\delta [S(\rho_{\pi}||\rho_*)-S(\rho||\rho_*)].
\end{equation}
This decomposition reveals a refined structure underlying the second law of quantum thermodynamics. In the equilibrium regime (i.e., when $\rho=\rho_{\pi}$), note that the passive component completely determines entropy production, while the non-passive term contributes exclusively in nonequilibrium conditions.

We can rewrite the second law in a generalized Clausius form by considering thermal maps, i.e., quantum evolutions for which the fixed point is a Gibbs state at some equilibrium temperature $T_e$:
\begin{equation}
\rho_*=\rho_e=\frac{e^{-H/k_BT_e}}{\text{tr}[e^{-H/k_BT_e}]} ,
\end{equation}
where $T_e$ is provided by Eq.~(\ref{TGibbs}). 
In this case, we obtain
\begin{equation}
\delta\Sigma=dS+\frac{\delta Q_{e}}{T_e}\geq0,
\end{equation}
where 
\begin{equation}
 \delta Q_e\equiv -dU+\text{tr}[\rho_e d H]   
\end{equation}
defines an effective heat associated with the environment. The passive and non-passive parts of the entropy production are given by 
\begin{equation}
 \delta\Sigma_{\pi}=-\delta S(\rho_{\pi}||\rho_e)=dS+\frac{\delta Q_{e \pi }}{T_e}  
\end{equation}
and
\begin{equation}
 \delta\Sigma_{n\pi}=-\frac{d\mathcal{E}}{T_e}, 
\end{equation}
where we identify $S(\rho_{\pi}||\rho_e)$ as the classical relative entropy~\cite{Sone2021}, with $\delta Q_{e \pi }$ representing the passive part of the effective heat. This reduces to $\delta Q_{e \pi}=- \delta Q$ for time-independent Hamiltonians, with $\delta\Sigma_{\pi}$ then leading to the classical Clausius inequality ($\delta\Sigma_{n\pi}=0$ in the equilibrium regime).

\section{Comparison with previous formulations}

In the standard  quantum formulation \cite{Alicki1979}, heat and work are defined by changes in $\rho$ and $H$, respectively:
\begin{equation}
 \delta \mathcal{Q}\equiv \text{tr}[d\rho\, H]   
\end{equation}
and
\begin{equation}
\delta \mathcal{W} \equiv \text{tr}[\rho \, dH],
\end{equation}
such that $dU=\delta\mathcal{Q}+\delta\mathcal{W}$. Since $\delta \mathcal{Q}$ is not invariant under passive transformation, i.e., $\delta \mathcal{Q}\neq \delta \mathcal{Q}_{\pi}$, the conventional heat is not necessarily connected with $dS$. In the entropy-based framework \cite{Alipour2022}, an additional work $\mathcal{W^{*}}$ narrows the connection between heat and entropy variation: 
\begin{equation}
 \delta\mathbbm{Q} \equiv\delta\mathcal{Q}-\delta\mathcal{W^{*}}   
\end{equation}
and 
\begin{equation}
\delta\mathbbm{W}\equiv \delta\mathcal{W}+\delta\mathcal{W^{*}},    
\end{equation}
where 
\begin{equation}
\delta \mathcal{W}^*\equiv \text{tr}[\delta\rho^{ep} H]    
\end{equation}
with 
\begin{equation}
\delta \rho^{ep}=\sum_n r_n d(\ket{r_n}\bra{r_n})    
\end{equation}
representing the change in $\rho$
due to eigenprojector variations. Indeed, we can write $\delta\mathbbm{Q}=\text{tr}[\delta \rho^{ev}H]$ and $dS=\text{tr}[\delta \rho^{ev}\sigma]$, where $\delta \rho^{ev}=\sum_n dr_n \ket{r_n}\bra{r_n}$ denotes the change in $\rho$ due to eigenvalue variations (note that $d\rho=\delta\rho^{ev}+\delta\rho^{ep}$). However, although both $\delta\mathbbm{Q}$ and $dS$ depend on $\delta\rho^{ev}$, we have $\delta\mathbbm{Q}\neq \delta\mathbbm{Q} _{\pi}$ and, consequently, the entropy-based heat is also not completely linked to entropy variation. Since $\mathcal{W}^{*}$ and $\mathcal{E}$ are purely non-passive quantities, we have the following connections among the three formulations:
\begin{equation}
\delta Q=\delta \mathcal{Q}_{\pi}=\delta \mathbbm{Q}_{\pi}   
\end{equation}
and 
\begin{equation}
\delta W_{\pi}=\delta \mathcal{W}_{\pi}=\delta \mathbbm{W}_{\pi}. 
\end{equation}
 
 There is also an operational formulation involving ergotropy, where energy variation is divided into three parts \cite{Binder2015}: 
 \begin{equation}
 \Delta U=Q_{op}+W_{ad}+\Delta\mathcal{E}   
 \end{equation}
 for a general and finite quantum process $(\rho_i,H_i)\rightarrow (\rho_f,H_f)$, where 
 \begin{equation}
 Q_{op}\equiv \text{tr}[\pi_m H_i]-\text{tr}[\rho_{i\pi} H_i]    
 \end{equation}
 and 
 \begin{equation}
 W_{ad}\equiv\text{tr}[\rho _{f\pi} H_f]-\text{tr}[\pi_m H_i]    
 \end{equation}
 define the operational heat and the adiabatic work, respectively, with
 \begin{equation}
 \pi_m\equiv \sum_{n} r_{nf}\ket{\varepsilon_{ni}}\bra{\varepsilon_{ni}}    
 \end{equation}
 corresponding to an auxiliary state. Note that 
 \begin{equation}
\Delta U_{\pi}=Q+W_{\pi}=Q_{op}+W_{ad}.
 \end{equation}
 In particular, we have $Q=Q_{op}$ and $W_{\pi}=W_{ad}=0$ when $d H=0$. Thus, assuming $W_{ad}+\Delta\mathcal{E}$ as the total work, the operational and the ergotropy-based formulations are equivalent for time-independent Hamiltonians. However, the equivalence between the two formulations fails for time-dependent Hamiltonians. For example, we have $Q_{op}=0$,  $\forall \,H(t)$, such that the initial Hamiltonian $H(0)=0$. 

\section{Qubit thermodynamics}

Let us consider an arbitrary qubit system, where
\begin{equation}
\rho=(\mathbbm{I}+\vec{r}\cdot \vec{\sigma})/2
\end{equation}
and 
\begin{equation}
H=-\vec{h}\cdot \vec{\sigma} ,   
\end{equation}
with $\vec{r}=(x,y,z)$ representing the Bloch vector, $\vec{\sigma}=(\sigma_x,\sigma_y,\sigma_z)$ the Pauli operators, and $\vec{h}=(h_x,h_y,h_z)$ the local field. In this scenario \cite{Vallejo2020, Choquehuanca2023,Vallejo2021}, we have 
\begin{equation}
\delta \mathcal{Q}=-\vec{h}\cdot d\vec{r}    
\end{equation}
and 
\begin{equation}
\delta \mathcal{W}=-\vec{r}\cdot d\vec{h}    
\end{equation}
for the conventional formulation, while 
\begin{equation}
\delta \mathbbm{Q}=(U/r)dr   
\end{equation}
and 
\begin{equation}
\delta \mathbbm{W}=r d(U/r) ,
\end{equation}
with 
\begin{equation}
U=-\vec{h}\cdot \vec{r},   
\end{equation}
in the entropy-based approach. For the ergotropy-based framework, from Eq.~(\ref{heatwork}), we then obtain
\begin{equation}\label{eq:Qergotropy}
\delta Q=-h dr
\end{equation}
and
\begin{equation}
\delta W=-rdh+d\mathcal{E},   
\end{equation}
where 
\begin{equation}
 \mathcal{E}=U+hr.   
\end{equation}
The expressions for the temperature of a qubit have been obtained for the conventional and entropy-based formulations through the derivative of the von Neumann entropy with respect to energy in a zero work
process \cite{Vallejo2020,Vallejo2021}: 
\begin{equation}
\mathcal{T}=\frac{h^2 r}{k_B (\vec{h}\cdot\vec{r})\text{tanh}^{-1}r}   
\end{equation}
and
\begin{equation}
 \mathbbm{T}=\frac{\vec{h}\cdot\vec{r}}{k_B r\text{tanh}^{-1}r},   
\end{equation}
with
the conventional temperature $\mathcal{T}$ compatible with the temperature defined in Eq.~(\ref{temperature}). From Eq.~(\ref{ptemperature}), we obtain the ergotropy-based temperature
\begin{equation}\label{eq:ergoT}
T=\frac{h}{k_B \text{tanh}^{-1}r}.
\end{equation}
Note that 
\begin{equation}
T=\mathcal{T}_{\pi}=\mathbbm{T}_{\pi}.    
\end{equation}
Furthermore, since 
\begin{equation}
dS=-k_B\text{tanh}^{-1}rdr,    
\end{equation}
we can write
\begin{equation}
 dS=\frac{\delta Q}{T}.   
\end{equation}
Remarkably, the ergotropy-based formulation applied to a qubit system leads to an expression 
that resembles the well-known classical relation between entropy and reversible heat.
\section{Applications}
\subsection{Qubit under generalized amplitude damping}

\begin{figure}[h!]
\centering
\includegraphics[width=0.97\columnwidth]{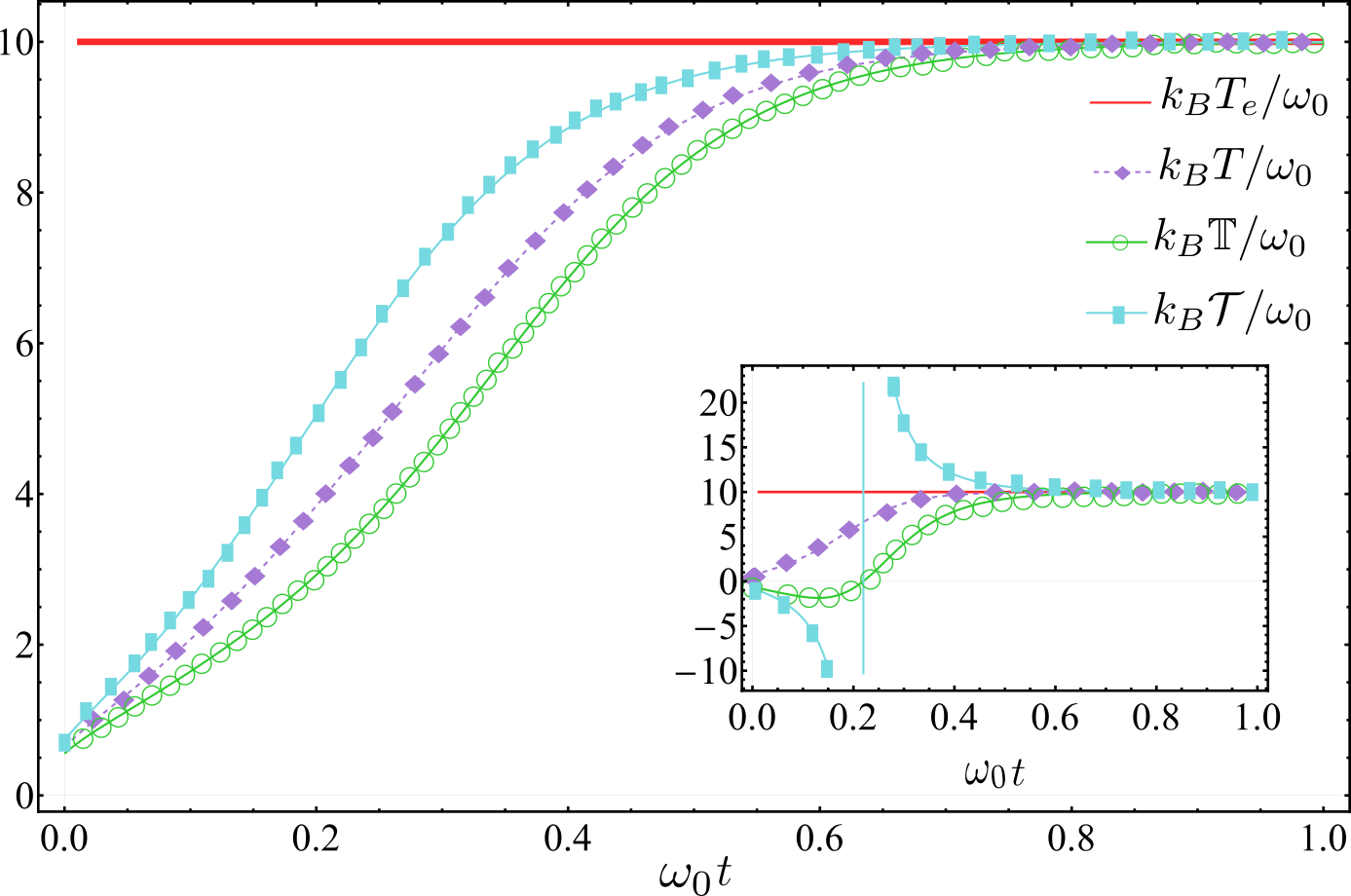}
\caption{(Color online) Dimensionless temperatures $k_BT_e/\omega_0$,  $k_BT/\omega_0$, $k_B\mathbbm{T}/\omega_0$, and $k_B\mathcal{T}/\omega_0$ as functions of the dimensionless time $\omega_0t$  for a qubit under a Markovian GAD process with $\vec{r}_{-}(0)=(0.45,\,0.00,\,-0.80)$. Inset: Same functions for the initial state $\vec{r}_{+}(0)=(0.45,\,0.00,\,0.80)$. We have used $\gamma_0=1$.}
\label{fig:GAD}
\end{figure}

Consider the Markovian quantum master equation for a generalized amplitude damping (GAD) process (we adopt $\hbar=1$), 
\begin{equation}
\frac{d\rho(t)}{dt}=-i[H(t),\rho(t)]\,+\mathcal{D}^-[\rho(t)]\,+\mathcal{D}^+[\rho(t)],    
\end{equation}
which describes a qubit interacting with a bosonic thermal reservoir at finite temperature $T_e$~\cite{Alipour2016,Yang2024,camati2019,cherian2019,Breuer2002,zeng2024}, where 
\begin{equation}
\mathcal{D}^{\mp}[\rho(t)]=\gamma^{\mp}\,(\sigma^{\mp}\rho(t)\sigma^{\pm} - \frac{1}{2}\{\sigma^{\pm}\sigma^{\mp},\rho(t)\})    
\end{equation}
runs the emission/absorption process, with
$\gamma^- =\gamma_0(N+1)$ and $\gamma^+=\gamma_0 N$,  $\sigma^\pm=\sigma_x \, \pm \, i\sigma_y$ are the ladder operators, $N=(e^{\beta_e  \omega_0}-1)^{-1}$ is the Planck distribution at frequency $\omega_0$, and $\beta_e=(k_B T_e)^{-1}$ is the inverse temperature of the environment. Assuming $\vec{h}=(0,0,-\omega_0/2)$ and $k_{B}T_e=10\omega_0$, we numerically solve the master equation for the qubit initially prepared in the mixed state $\vec{r}_{\pm}(0)=(0.45,\,0.00,\,\pm0.80)$ (upper and lower hemispheres of the Bloch sphere). Fig. \ref{fig:GAD} illustrates the dynamical behaviors of the three different temperature definitions ($\mathcal{T}$, $\mathbbm{T}$, and $T$). All temperature quantifiers converge to the environment temperature at long times. Observe that the conventional temperature exhibits nonanalytical behavior at $\omega_0 t \approx0.2195$ for the initial state $\vec{r}_+(0)$. At this time, the state exhibits the Bloch vector component $z=0$. Note also that, as shown before, the ergotropy-based temperature is always positive.

\subsection{Qubit under phase damping}
\begin{figure}[h!]
\centering
\includegraphics[width=0.97\columnwidth]{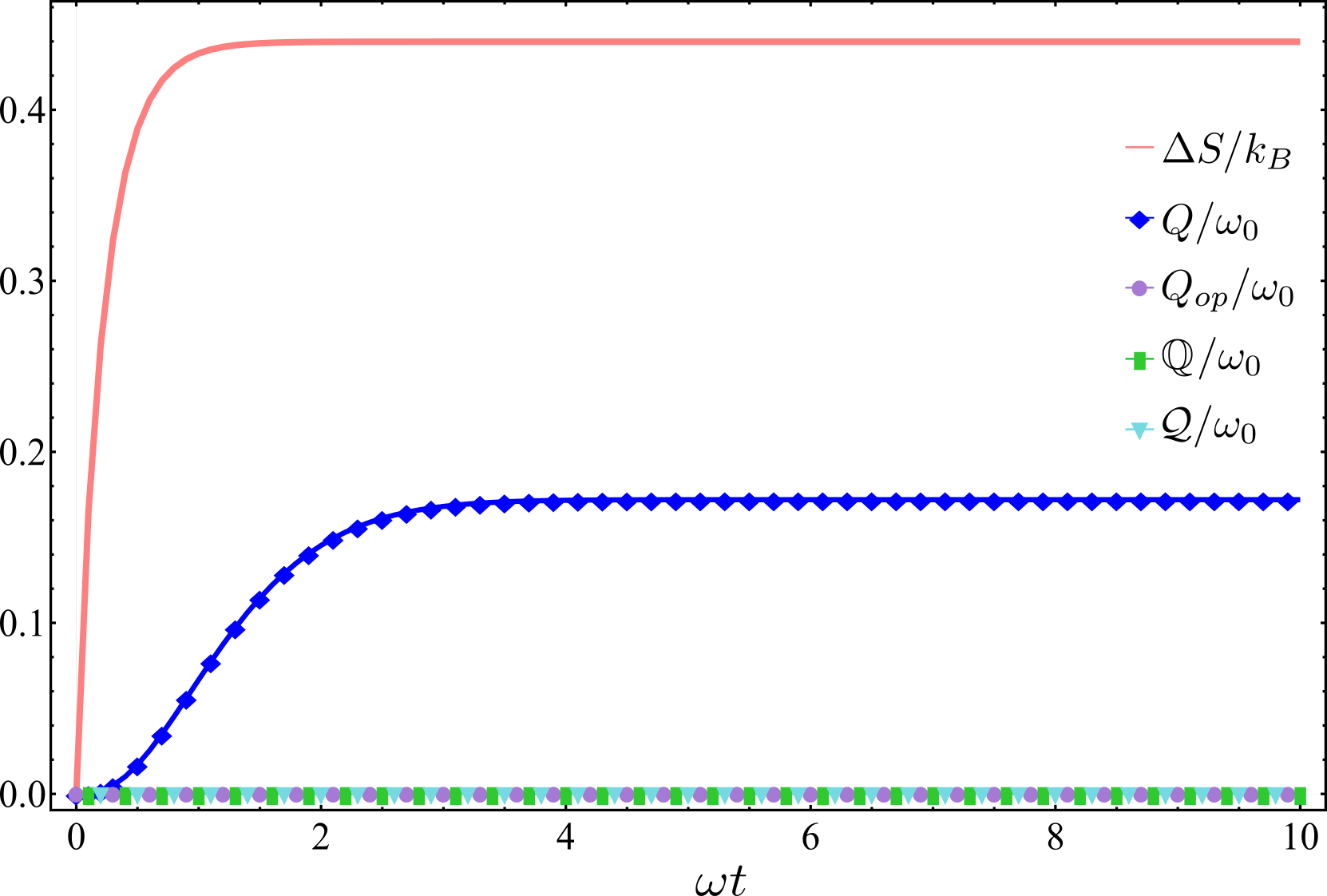}
\caption{(Color online) Dimensionless heats $Q/\omega_0$, $Q_{op}/\omega_0$, $\mathbbm{Q}/\omega_0$, $\mathcal{Q}/\omega_0$, and entropy variation $\Delta S/k_B$  as functions of the dimensionless time $\omega t$ for a qubit under a Markovian PD process with $\vec{r}(0)=(0.5,\,0.7,\, 0.0)$ and $\gamma=\omega$.}
\label{fig:PDM}
\end{figure}

Let us consider a qubit under a phase damping (PD) dynamics~\cite{Choquehuanca2023,Marcantoni2017,Garcia2022,Yang2024},
\begin{equation}\label{PD}
\frac{d\rho(t)}{dt}=-i[H(t),\rho(t)]\,+ \gamma(t)(\sigma_z \rho(t)\sigma_z - \rho(t)) ,   
\end{equation}
assuming a time-dependent Hamiltonian, with $\vec{h}(t)=[0,0,-\omega_0(1-\cos{\omega t})/2]$~\cite{scopa2019}, a time-independent decoherence
rate $\gamma(t)=\gamma$ (i.e., the Markovian regime), and an arbitrary initial state $\vec{r}(0)=(x_0,y_0,z_0)$. In this case, the solution is given by $\vec{r}(t)=[x(t),y(t),z(t)]$ with
\begin{equation}
    x(t)=e^{-2t\,\gamma} \left( x_0 \cos{\alpha} - y_0 \sin{\alpha} \right),
\end{equation}
\begin{equation}
 y(t)=e^{-2t\, \gamma} \left( y_0 \cos{\alpha} + x_0 \sin{\alpha} \right),   
\end{equation}
\begin{equation}
    z(t)=z_0,
\end{equation}
where 
\begin{equation}
\alpha=\omega_0( \omega t -\sin \omega t)/\omega.    
\end{equation}
To demonstrate the stronger connection of the ergotropy-based heat with entropy over other heat formulations, we consider the evolution in the $xy$-plane of the Bloch sphere ($z_0=0$), for which only the ergotropy-based heat is non-vanishing and monotonically related to von Neumann entropy, as shown in Fig.~\ref{fig:PDM}. In particular, $Q(t)$ resembles the behavior of  classical heat in a reversible process~\cite{sears1975,Tahir2020,prasad2016,cheng2006}. It can be viewed as an informational heat, quantifying decoherence through 
\begin{equation}
\Delta S(t)=\int \frac{\delta Q(t)}{T(t)},   
\end{equation}
which increases as coherence is lost. Meanwhile, $T(t)$ acts as an internal parameter controlling the degradation of quantum information as a function of time.

\subsection{Quantifying non-Markovianity via heat}

We now explore a scenario in which a qubit is coupled to a non-Markovian PD noise governed by Eq.~(\ref{PD})~\cite{Haikka2011, Breuer2002}. We assume a zero-temperature bosonic reservoir with an Ohmic-like spectral density, where the time-dependent decoherence rate is 
\begin{equation}
\gamma(t,s)=\left[1+(\omega_c t)^2\right]^{-s/2}\, \Gamma_e[s]\, \sin{[s\,\arctan(\omega_c t)]}    
\end{equation}
with $\Gamma_e[x]$ denoting the Euler gamma function, $\omega_c$ the reservoir cutoff frequency, and $s\geq 0$ the ohmicity parameter. Depending on the value of $s$, the model can exhibit either Markovian or non-Markovian dynamics \cite{Dhar2015,Haseli2015,Addis2014,Haikka2013,Dakir2025}, with $0\leq s\leq 2$ and  $s > 2$ corresponding to the Markovian and non-Markovian regimes, respectively. The solution for a time-independent Hamiltonian, with  $\vec{h}=(0,0,-\omega_0)$, and an arbitrary initial state $\vec{r}(0)=(x_0,y_0,z_0)$ is given by $\vec{r}(t)=[x_0\, \Gamma(t),y_0\, \Gamma(t),z_0]$,
where $\Gamma(t)=\exp{(\int^t_0 \gamma(t)dt)}$ \cite{Choquehuanca2023}. By expressing the initial state in spherical coordinates, $\vec{r}(0)=(r_0\sin{\theta_0}\cos{\phi_0}, \,r_0\sin{\theta_0} \sin{\phi_0},\, r_0\cos{\theta_0})$, we find 
\begin{equation}
Q(t, r_0,\theta_0)=-\omega_0\,r_0\,([\cos^2{\theta_0}+\Gamma^2(t)\, \sin^2{\theta_0})]^{1/2}-1)    
\end{equation}
for the ergotropy-based heat. Since $Q$ is monotonically related to the entropy for an arbitrary qubit state, we can use $Q$ to characterize non-Markovianity for unital maps~\cite{Haseli2015,Choquehuanca2023}. 
More specifically, we observe that, such as entropy, $Q$ displays a monotonic decreasing behavior as a function of time for unital CPTP divisible maps~\cite{Breuer:09,Rivas:10,Vacchini:11} (see also Ref.~\cite{Huang:21} for a recent correlation-based non-Markovianity measure). Non-Markovianity can then be captured by a violation of this monotonic behavior, which will occur depending on the value of the ohmicity parameter $s$.
In this direction, we adopt the generalized approach recently proposed in Ref.~\cite{Choquehuanca2023} and compute the corresponding non-Markovianity measure 
\begin{equation}
N_{Q}=\max_{\rho(0)} \sum^{}_{i}{\abs{Q(a_i,r_0,\theta_0)-Q(b_i,r_0,\theta_0)}},    
\end{equation}
being $[a_i,b_i]$ the set of time intervals for which  $\gamma(t,s)\leq0$, with $i=1,2,3,...$ labeling the number of such intervals for a given range of $s$. Specifically, for $s\leq 2$, there are no intervals where $\gamma$ becomes negative. For $2<s\leq6$, a single negative interval emerges $(i=1)$, and for $s>6$, the number of such intervals increases $(i=2,3,...)$. Here, we focus on the case of a single interval (see Refs.~\cite{Addis2014, Haikka2013}). We find that the optimal initial state in the definition of $N_Q$ is a pure state $\vec{r}_{max}(0)=(\sin\phi_0,\cos\phi_0,0)$,  with $0\leq\phi_0\leq2\pi$. Consequently,
\begin{equation}\label{nq}
N_{Q}=\omega_0\sum^{}_{i}{\abs{\Gamma(a_i)-\Gamma(b_i)}}.  
\end{equation}
This expression captures the memory effects as the widely employed trace distance-based measure~\cite{Breuer:09}, given by
\begin{equation}
N_D=\max_{\vec{r}_1(0), \vec{r}_2(0)} \sum_i |D(\vec{r}_1(a_i),\vec{r}_2(a_i))- D(\vec{r}_1(b_i),\vec{r}_2(b_i))|,    
\end{equation}
where $D(\vec{r}_1,\vec{r}_2)=|\vec{r_1}-\vec{r_2}|/2$ defines the trace distance between the states $\vec{r}_2$ and $\vec{r}_1$. In fact, the optimal pair of states in $N_{D}$ corresponds to $\vec{r}_2(0)=-\vec{r}_1(0)=(1,0,0)$ \cite{wissmann2012,breuer2016,lorenzo2013}, which leads to
\begin{equation}
 N_Q=\omega_0N_D.   
\end{equation}
Notably, the expression in Eq.(\ref{nq}) also holds for the coherence-based measure \cite{Choquehuanca2023}.

\begin{figure}. 
\centering
\includegraphics[width=0.97\columnwidth]{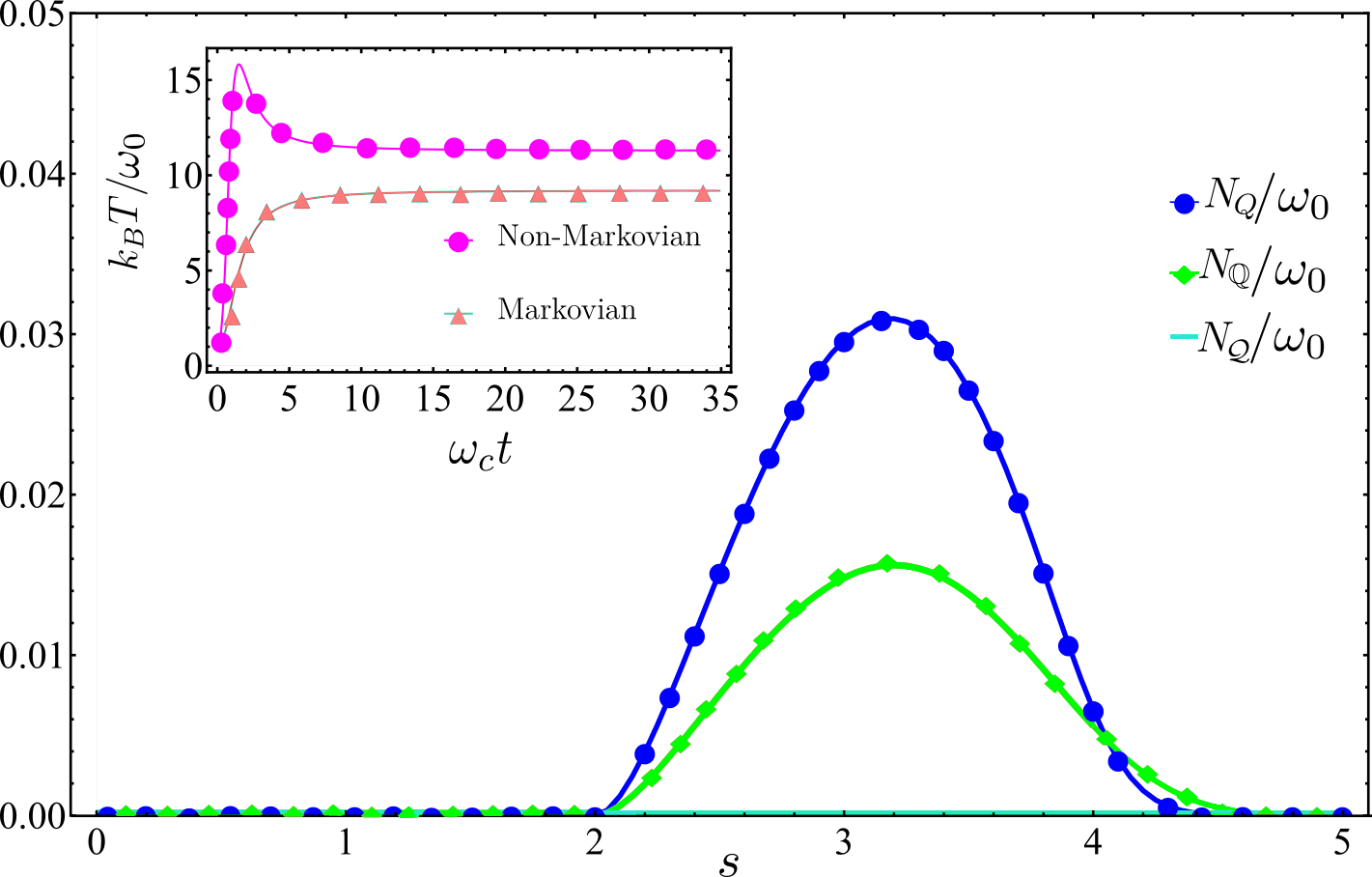}
\caption{(Color online) Dimensionless heat-based non-Markovianity quantifiers $N_Q/\omega_0$, $N_{\mathbbm{Q}}/\omega_0$, and $N_{\mathcal{Q}}/\omega_0$ as functions of the ohmicity parameter $s$. Inset: Dimensionless ergotropy-based temperature $T$ as a function of $\omega_c t$ for $|\vec{r}(0)|=0.8$.}
\label{fig:PDNM}
\end{figure}

In Fig.~\ref{fig:PDNM}, we compare $N_Q/\omega_0$ with the heat-based alternatives $N_\mathcal{Q}/\omega_0$ and $N_\mathbbm{Q}/\omega_0$. Note that $N_\mathcal{Q}/\omega_0=0$ for all $s$, which means that $N_\mathcal{Q}$ is indeed unsuitable as a non-Markovianity measure. Concerning $N_\mathbbm{Q}/\omega_0$, the quantification provided is, on average, numerically less pronounced than $N_{Q}/\omega_0$. Specifically, the maximum non-Markovianity is observed  at $s=3.2$, yielding $N_{\mathbbm{Q}}/\omega_0 \approx0.0156$ and $N_Q/\omega_0 \approx 0.0309$.  
Moreover, $N_\mathbbm{Q}$ is a more restricted measure, only valid for an energy sign-preserving unital map~\cite{Choquehuanca2023}. In the inset of Fig.~\ref{fig:PDNM} one can see how the definition of the ergotropy-based temperature, Eq.~\eqref{eq:ergoT}, is applicable as a witness of non-Markovianity through its non-monotonic behavior over time. The figure highlights both Markovian $(s=2)$ and non-Markovian $(s=3.2)$ regimes. For this analysis, the system is initialized in a mixed state $|\vec{r}(0)|=0.8$. Notice that $T$ successfully captures non-Markovianity within the $xy$-plane of the Bloch sphere, unlike the alternative temperature definitions $\mathcal{T}(t)$ and $\mathbb{T}(t)$, which fail to detect such behavior.

\section{Conclusions}

We have introduced an ergotropy-based formulation of quantum thermodynamics. This framework allowed for a direct relationship between heat and von Neumann entropy, which is stronger than the connection found in previous approaches. This is based on the invariance of the ergotropy-based heat under passive state transformations. In this scenario, average heat can then be used as a general measure of non-Markovianity for unital maps. Moreover, by defining the out-of-equilibrium temperature in an ergotropy-based formulation, we can achieve a positive-semidefinite temperature. This means that, even working in a nonequilibrium context, temperature will follow a simple description  typical of equilibrium states, with non-negative values throughout the dynamics. Concerning work, we have obtained that average work is provided by the ergotropy variation and an extra passive work contribution, which can be induced by either a controllable parameter of the system or even by the interaction with the environment. As future perspectives, we intend to look at the efficiency of quantum thermal machines in the ergotropy-based scenario, both by theoretical and experimental proposals. In addition, we can also explore the ergotropy-based framework in terms of a resource theory for energy extraction in open quantum systems (see, e.g., Ref.~\cite{Alhambra:19}). We leave such topics for future developments.

{\section*{Acknowledgments}}

J.M.Z.C. acknowledges Conselho Nacional de Desenvolvimento Cient\'{\i}fico e Tecnol\'ogico (CNPq) for financial support. 
M.S.S. is supported by Conselho Nacional de Desenvolvimento Cient\'{\i}fico e Tecnol\'ogico (CNPq) under the grant number 303836/2024-5. 
This research is also supported in part by Coordena\c{c}\~ao de Aperfei\c{c}oamento de Pessoal de N\'{\i}vel Superior (CAPES) (Finance Code 001) 
and by the Brazilian National Institute for Science and Technology of Quantum Information (INCT-IQ).


%

\end{document}